\documentclass[12pt,reqno]{amsart}

\usepackage{amsaddr}
\usepackage[authoryear,round]{natbib}
\usepackage{float}
\usepackage[hidelinks, linktocpage=true]{hyperref}
\usepackage[pdftex]{graphicx}
\usepackage{caption}
\usepackage[a4paper, margin=2.3cm]{geometry}
\usepackage{orcidlink}

\begin{document}

\title{The Zero Degree of Freedom Non-Central Chi Squared Distribution for Ensemble Postprocessing}

\author{J\"{u}rgen Gro{\ss}{$^{1}$}\orcidlink{0000-0002-3861-4708}}
\address{Institute for Mathematics and Applied Informatics, University of Hildesheim, Germany}
\email{$^1$juergen.gross@uni-hildesheim.de}
\author{Annette M\"{o}ller{$^{2}$}\orcidlink{0000-0001-9386-1691}}
\address{Faculty of Business Administration and Economics,  Bielefeld University, Germany}
\email{$^{2}$annette.moeller@uni-bielefeld.de}


\date{}

\begin{abstract} In this note the use of the zero degree non-central chi squared distribution
as predictive distribution for ensemble postprocessing is investigated. It has a point mass at zero by definition, and is thus particularly suited for postprocessing weather variables naturally exhibiting large numbers of zeros, such as precipitation, solar radiation or lightnings. Due to the properties of the distribution no additional truncation or censoring is required to obtain a positive probability at zero. The presented study investigates its performance
compared to that of the censored generalized extreme value distribution and the censored and shifted gamma distribution for postprocessing 24h accumulated precipitation using an EMOS (ensemble model output statistics) approach with a rolling training period. The obtained results support the conclusion that it
serves well as a predictive distribution in postprocessing precipitation and thus may also be considered in future analyses of other weather variables having substantial zero observations.
\end{abstract}

\maketitle

\markboth{J.~Gro{\ss} \& A.~M\"{o}ller}{Zero Degree of Freedom Distribution}

\section{Introduction}\label{sec:intro}

Forecasting of weather variables is usually based on numerical weather prediction (NWP) models. Such models are build from a set of differential equations to represent the dynamical physics of the atmosphere. Then, the equations are discretized in space and time, and the current state of the atmosphere is evolved forward in time. The obtained solutions strongly depend on initial conditions and model formulations, thus NWP models suffer from various sources of uncertainties. These uncertainties in model formulations and initializiation may be addressed by using ensemble prediction systems (EPS) consisting of multiple runs of the NWP model. Each time variations in the  parameterizations and initial as well as boundary conditions are employed \citep{GneitingRaftery2005, LeutbecherPalmer2008}.

A forecast ensemble can be viewed as a probabilistic forecast that allows to assess forecast uncertainty \citep{Palmer2002}. Nonetheless, NWP ensembles typically exhibit forecast biases and dispersion errors, which may then be addressed by statistical postprocessing methods \citep{GneitingRaftery2005, gneiting2014probabilistic, WilksHamill2007}. Such methods aim at improving calibration and forecast skill and provide full predictive probability distributions.

Many of the proposed statistical postprocessing models are extensions and modifications of two generic state-of-the-art models, namely the Ensemble Model Output Statistics approach (EMOS, see \citealp{gneiting2005calibrated}) and the Bayesian Model Averaging (BMA, see \citealp{Raftery&2005}). Besides adapting the methods to the requirements of different weather variables, there is also a growing interest in extending the basic models to be able to explicitly incorporate multivariate dependence structures. For example, \citet{moller2013multivariate, baran2015joint} have applied BMA jointly with a Gaussian copula approach to obtain a multivariate predictive distribution for different weather variables jointly. \citet{jobst2023d} considered extensions of the zero-truncated ensemble model approach by employing D-vine copulas. See also \citet{jobst2023d2} for related work.
Apart from modelling dependencies among weather variables, incorporation of dependencies between consecutive time points recently gained attention as well. For example \citet{moller2016probabilistic, moller2020probabilistic} have used a times series EMOS approach for temperature forecasting. This approach was further extended in \citet{jobstetal2024}.

In this note we are concerned with probabilistic forecasting of accumulated precipitation based on past observations as well as past and actual ensemble forecasts with an EMOS approach utilizing the zero degree non-central chi squared distribution. The aim of this work is to conduct a benchmark comparison with postprocessing models based on other distributions suitable for the task of precipitation forecasting. However, future research will also be concerned with the development of multivariate postprocessing methods making use of the this distribution.

An extension of BMA to probabilistic quantitative precipitation forecasting has been given by \citet{sloughter2007probabilistic}, where a mixture of a discrete component at zero and a gamma distribution is used. Since then, further applications of Bayesian model averaging for precipitation forecasting have been developed, as an example consider \citet{saedi2020performance}. BMA methods are available in {\sf R} with the package {\tt ensembleBMA} \citep{RPensembleBMA}. A Bayesian ensemble postprocessingg approach based on copula
functions is considered by \citet{khajehei2017towards}. In connection with EMOS, \citet{scheuerer2014probabilistic} introduced the  generalized extreme value distribution left-censored in zero (GEV0). Moreover, \citet{scheuerer2015statistical} and also \citet{baran2016censored} specified the censored, shifted gamma distributions (CSG0) for this task.  Both, GEV0 and CSG0, are available with the {\sf R} package {\tt ensmbleMOS}, see \citet{RPensembleMOS}. A comparison of
BMA with EMOS methods for precipitation forecasting is given by  \citet{javanshiri2021comparison}. Further censored or truncated distributions had also been considered for precipitation forecasts. See \citet{stauffer2017ensemble} for a left-censored power-transformed logistic response distribution or \citet{stauffer2017spatio} and
\citet{schlosser2019distributional} for applications of a zero-censored Gaussian distribution.

The above overview suggests that while there are several state-of-the-art distributions established for a variable such as precipitation with naturally occurring zero observations, none of them is able to outperform the others in every aspect. The performance strongly depends on characteristics of the observation and ensemble data at hand.
The difficulty in accurately predicting zero as well as extreme precipitation events and defining an adequate postprocessing model is the motivation for the introduction of a novel alternative predictive distribution and a more thorough exploration of its properties compared to the established distributions. We hope that this study provides further insight for researchers and forecasters when it comes to choosing an appropriate distribution for probabilistic precipitation forecasting.

In the following, we consider GEV0 and CSG0 as our benchmark distributions within the classical EMOS approach, and propose the zero degree of freedom non-central chi square squared distribution (Chi0) as a further alternative. Since it has by definition a point mass at zero, no truncation or censoring is required in order to accommodate weather variables exhibiting larger amounts of zero observations. The non-centrality parameter controls the point mass at zero as well as the distributional shape. For the sake of  flexibility it is convenient to introduce an additional scale parameter, see \citet{siegel1985modelling}.
These additional features make the distribution an interesting candidate for probabilistic precipitation forecasting, but is also worth investigating for other variables sharing some of the difficult properties of precipitation, such as e.g. solar radiation or lightnings.

\section{Zero Degree of Freedom Non-Central Chi Squared Distribution}

The cumulative distribution function (CDF) of the $\chi_{\nu}^2(\lambda)$ distribution  with $\nu >0$ degrees of freedom and non-centrality parameter $\lambda \geq 0$ may be written as
\begin{equation}\label{E1}
F(x; \nu, \lambda) = \sum_{j=0}^{\infty}  p(j) \, F(x; \nu + 2j; 0), \quad x \geq 0\; ,
\end{equation}
where
\begin{equation}
p(j) = \frac{(\lambda/2)^{j}}{j!}e^{- \lambda/2}, \quad j =0,1,2,\ldots
\end{equation}
is the probability mass function of the Poisson distribution with parameter $\lambda/2$, see
\citet{johnson1995continuous}. When considering  the case $\nu=0$ and letting $F(x; 0; 0) =1$ for all $x\geq 0$, the function $F_{0}(x; \lambda) := F(x; 0, \lambda)$ from \ref{E1} defines the so-called  non-central $\chi^2$ distribution with zero degrees of freedom denoted by $\chi_{0}^2(\lambda)$. It has a point probability mass of $e^{-\lambda/2}$ at $x= 0$ and thus is a mixture of
$0$, $\chi_{2}^2(0)$, $\chi^2_{4}(0)$, \ldots  with Poisson weights $p(j)$. This distribution has been defined by \citet{torgersen1972supplementary} and apparently independently by \citet{siegel1979noncentral} who derived a number of its properties. \citet{hjort1988eccentric} calls it the eccentric part of the non-central $\chi^2$ distribution when introducing a decomposition of this distribution. \citet{jones1987relationship} discusses the relationship with the Poisson-Exponential model.
For the purpose of computing the CDF of the $\chi_{0}^{2}(\lambda)$ distribution, the representation (\ref{E1}) may be employed with $F$ replaced by $F_{0}$ as noted above. However, the CDF is already implemented with the function {\tt pchisq} in the statistical software package {\sf R} \citep{Rsoftware}.

As it seems, the first application of the $\chi_{0}^2(\lambda)$ distribution equipped with an additional scale parameter $\sigma>0$  to  meteorological data has been given by \citet{siegel1985modelling} who fits it to January snowfall in Seattle from 1906 to 1960. As a generalization, \cite{dunn2004occurrence} considers the so-called Tweedie distributions for modelling precipitation, see also \citet{hasan2011two}.
In this note we seize the suggestion by \citet{siegel1985modelling} and propose the scaled CDF
\begin{equation}F_{0}(x/\sigma; \lambda)\end{equation} as a further alternative predictive CDF in ensemble postprocessing of weather variables admitting non-negative values with an expected accumulation at 0 as it is e.g. the case with (accumulated) precipitation amounts. The scaled $\chi_{0}^2(\lambda)$ distribution has
\begin{equation}\label{E6}
    \mu = \sigma \lambda\quad\text{and}\quad
    \mu_{2} = 4\sigma^2 \lambda
\end{equation}
as (predictive) mean $\mu$ and variance $\mu_{2}$, see \citet{siegel1985modelling}.

\begin{figure}[bt]
\centering
\includegraphics[width=30pc,angle=0]{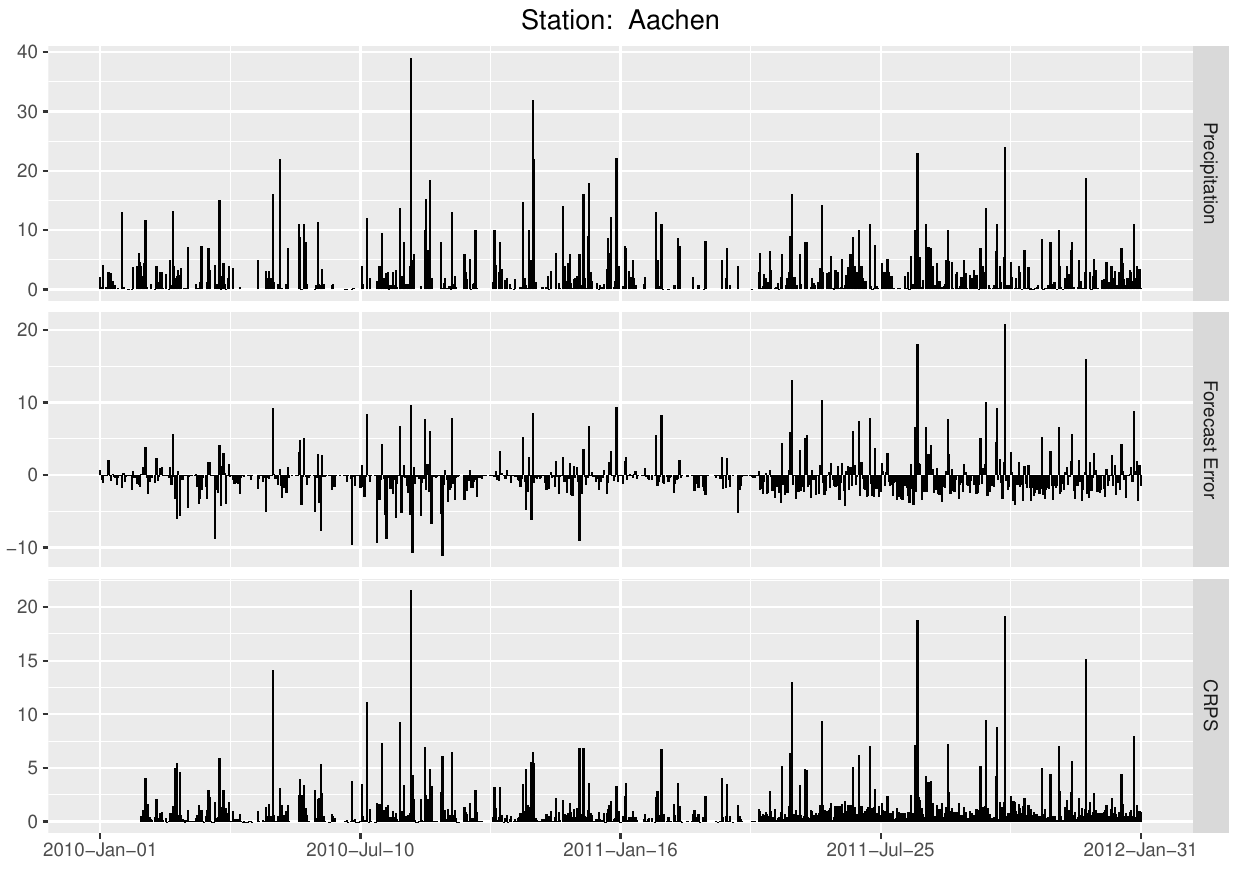}\\
  \caption{24 h accumulated precipitation amount for station Aachen as well as the observed amounts minus the corresponding ensemble means (forecast errors), and the CRPS values of the model Chi0}\label{F1}
\end{figure}

\section{Ensemble Postprocessing of Accumulated Precipitation}\label{sec:data}

We consider observations of 24h-accumulated precipitation amounts (in mm) from 01 January 2010 to 31 January 2012 (761 days) for 12 stations in Germany, see Table \ref{T1}. The stations were selected to represent low, medium and large precipitation values. The precipitation observations are matched with their corresponding $m=50$ exchangeable (24h ahead) member forecast European Center for Medium Range Weather Forecasts (ECMWF) ensemble forecasts, see \citet{molteni1996ecmwf,buizza2007new}. Our data is part of a larger data set also
employed by  \citet{hemri2014trends}.
All values (precipitation observations and forecasts) had been rounded off to one digit after the dot.
For the selected stations there is a total of $12\times 761 = 9132$ data points among which 366  were reported as missing (about 4\%). There only emerged one missing value pattern, involving the precipitation observation together with all its ensemble forecasts. In order to retain comparable consecutive days for all stations, the missing values had been imputed by chained equations as provided by the package {\tt mice} \citep{miceRpackage}, where only one set of imputed values had been  used. Prior and subsequent to imputation, there is a total of $1078$ data points admitting a precipitation forecast of $0$ among all $m=50$ ensemble members, implying a zero ensemble mean as well as zero ensemble variance (approximately 12\% of data points). To account for such a situation, \citet{gebetsberger2016tricks} consider a so-called 'split-approach' by introducing an indicator variable referring to the fraction of zero value ensemble members being greater than a predefined split level $\nu$. Here, we only consider split level $\nu=1$ (all ensemble members equal to zero) in which case the two parameters $\lambda$ and $\sigma$ of the considered predictive distribution  do not depend on the forecast ensemble at all, see Eq. (\ref{E3}).

\subsection{Forecast Verification}

The goal of probabilistic forecasting is to maximize sharpness of the predictive distribution subject to calibration, see e.g. \citet{GneitingRaftery2007}. Calibration is a joint property of the forecast and verifying  observation, is measures the statistical consistency of the probabilistic forecast and the associated observation. Contrary, sharpness is a property of the forecasts alone and refers to the spread of the predictive distribution.
There exist various tools to assess above mentioned properties for a given probabilistic forecast.

A popular approach is the use of proper scoring rules to assess the quality of the probabilistic forecast.
Such a scoring rule $s(F, y)$ assigns a real-valued score to a pair $(F,y)$ of a predictive distribution $F$ and verifying observation $y$.
When $F$ denotes the predictive CDF obtained from past values for prediction of the verifying observation $y$ a possible representation of the continuous ranked probability score (CRPS, \citealp{GneitingRaftery2007}) is given by
\begin{equation}\label{E4}
\text{CRPS}(F, y) = \int_{-\infty}^{y}
F(x)^2 \, \text{d} x + \int_{y}^{\infty} (1- F(x))^2 \, \text{d} x\; ,
\end{equation}
see also \citet[Sect. 8.5.1]{wilks2011statistical}. It measures calibration and sharpness simultaneously, where smaller values indicate better predictive performance. This makes the CRPS particularly suitable for forecast assessment.

For several predictive cumulative distribution functions there exist explicit CRPS formulas, most of which being implemented with the {\sf R} package {\tt scoringRules}, see \citet{jordan2017evaluating}. For $F_{0}(x/\sigma; \lambda)$ we evaluate formula (\ref{E4}) by numerical integration with the {\sf R} function {\tt integrate}. It is also worthwhile to compare probabilistic score values  with a corresponding score from the raw ensemble. The usual formula, see e.g. \citet{grimit2006continuous}, is given by
\begin{equation}
\text{CRPS}(F_{\text{Ens}}, y) =
\frac{1}{m} \sum_{i=1}^{m} |f_{i} - y| - \frac{1}{2 m^2}
\sum_{i=1}^{m}\sum_{j=1}^{m} |f_{i} - f_{j}|\; ,
\end{equation}
where $f_{i}$ denotes the forecast from the $i$th ensemble member.  It is implemented with the {\sf R} packages {\tt ensembleBMA} and {\tt ensembleMOS} \citep{RPensembleBMA, RPensembleMOS}.

In addition to scoring rules, visual tools such as the verification rank and the PIT histogram are often used to assess calibration.

For a continuous CDF $F$ and a variable $Y$ that is distributed according to $F$ it follows  that $F(Y) \sim \text{Unif}[0,1]$ (probability integral transform, PIT). Evaluating the predictive distribution $F$ at the respective verifying observation $y$ for all forecast cases and plotting the binned frequencies of the resulting PIT values allows a visual assessment of calibration \citep{Dawid1984, GneitingRaftery2007}. Departures from uniformity indicate different types of miscalibration.
The discrete counterpart used for assessing the calibration of the forecast ensemble is the verification rank histogram \citep{Talagrand1997}. The rank of the verifying observation within the ordered ensemble is obtained for each forecast case and the rank frequencies plotted.
For a calibrated $m$-member ensemble the possible ranks $\{1,\ldots, m+1\}$ all appear equally likely, indicating a discrete uniform distribution on above set of ranks.

\begin{center}
\begin{table}
\begin{tabular}{lrrrr}
& $\overline{\text{CRPS(GEV0)}}$ & $\overline{\text{CRPS(CSG0)}}$ & $\overline{\text{CRPS(Chi0)}}$ & $\overline{\text{Ens}}$\\
\hline
Artern & 0.9191 (25.2) & 1.0374 (152.6)  & \textbf{0.8046} (27.8) & 0.8784 (17.4)\\
Erfurt Bindersleben & 1.0858 (39.8)& 0.8887 (30.5) & \textbf{0.8731} (30.5)& 0.9710 (29.4)\\
Berlin Tegel & 0.9224 (65.2) & 1.7209 (685.9) & \textbf{0.7524} (20.2) & 0.8261 (17.9)\\
Bad Salzuflen & 0.9890 (33.3)& 0.9399 (31.7) & \textbf{0.8642} (14.4)& 0.9213 (31.5)\\
Lahr & 1.2162 (50,2) & \textbf{1.1091} (21.1) & 1.1332 (21.4)& 1.3531 (20.1)\\
Weiden & 1.0506 (34.5)& 0.9714 (29.1) & \textbf{0.9638} (36.9)& 1.0298 (28.6)\\
Hamburg Fuhlsbuettel & 1.0188 (18.6) & 1.0277 (33.2) & 1.001 (20.3)& \textbf{0.9867} (19.4)\\
Duesseldorf & 1.0725 (30.8) & \textbf{0.9655} (31.7) & 0.9962 (31.8)& 1.0730 (31.3)\\
Aachen & 1.3109 (21.7) & 1.1794 (20.4) & \textbf{1.1747} (21.6) & 1.3080 (19.8)\\
Oberstdorf & 1.9290 (39.5)& 1.7398 (42.1)& \textbf{1.7068} (42.4)& 1.8736 (20.3)\\
Brocken & 1.8527 (35.3)& \textbf{1.8034} (35.7) & 1.8051 (36.1)& 2.4934 (39.5)\\
Zugspitze & 2.3814 (33.9) & 2.3269 (33.6) & \textbf{2.2839} (30.1)& 2.7103 (39.5)\\
\hline
\end{tabular}
\caption{Average CRPS values of the three models as well as the raw ensemble over 731 verification days with largest occurring value in round brackets}\label{T1}
\end{table}
\end{center}

\subsection{Linking Distribution Parameters to Ensemble Statistics}

In order to define the respective EMOS model, distribution parameters need to be linked to suitable predictor variables.
Here, we link the two parameters $\lambda$ and $\sigma$ of the predictive CDF $F_{0}(x/\sigma; \lambda)$ to summary statistics of the ensemble forecasts by
\begin{equation}\label{E3}
\lambda = a^2 +  b^2 \, \overline{f} \quad \text{and}\quad  \sigma = c^2 + d^2 \, s_{f}\; ,
\end{equation}
where $\overline{f}= \frac{1}{m} \sum_{i=1}^{m} f_{i}$ is the ensemble mean and
$s_{f} = \left(\frac{1}{m-1} \sum_{i=1}^{m}  (f_{i} - \overline{f})^2\right)^{1/2}$ is the ensemble standard deviation. As can be seen from (\ref{E6}), mean and variance of the scaled $\chi_{0}^{2}(\lambda)$ distribution both depend on the two parameters $\lambda$ and $\sigma$. Some preliminary (but maybe not exhaustive) initial runs suggested that a separate mapping of the two parameters to ensemble mean and ensemble standard deviation as in (\ref{E3}) works well with respect to optimization. Moreover it had been seen  that $s_{f}$ can be expected to perform better than the ensemble mean difference (MD) as employed by \citet{scheuerer2014probabilistic}.

The parameters $a,b,c,d$ are squared in (\ref{E3}) to ensure non-negativity and are obtained from minimizing the average CRPS over a training period of 30 days. For optimization the "Nelder-Mead" method with the {\sf R}  function {\tt optimize} had been used with strictly positive starting values. In case of all ensemble members equal to zero one gets $\lambda= a^2$ and $\sigma= c^2$.

For optimization with respect to GEV0 and CSG0, the corresponding {\sf R} functions provided by the package {\tt ensembleMOS} were utilized both with the method "L-BFGS-B". They work much quicker than optimization together with numerical integration as required for the Chi0 model.

\begin{figure}[bt]
\centering
\includegraphics[width=30pc,angle=0]{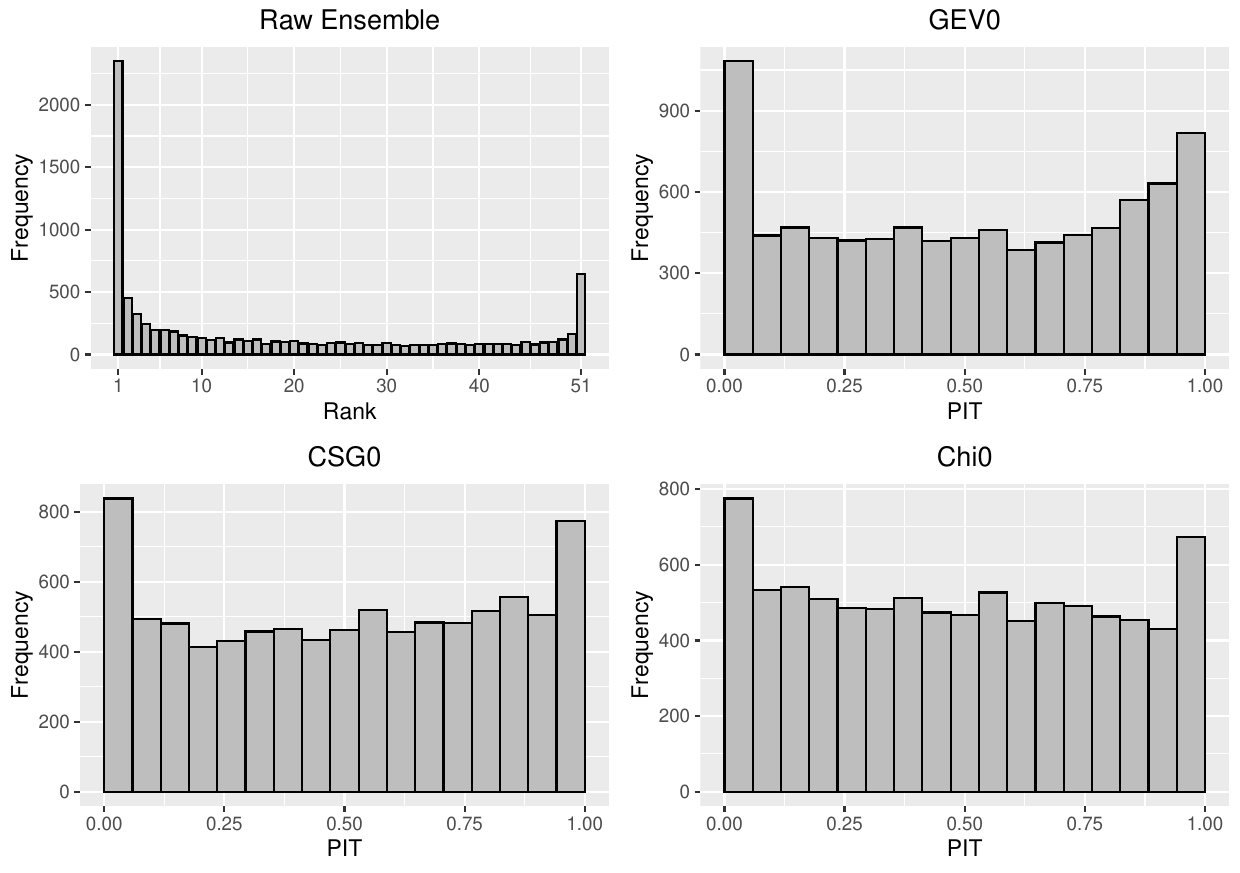}\\
  \caption{Rank verification bar plot of raw ensemble and PIT histograms of the three models, 12 stations and $12 \times 713 = 8772$ verification days}\label{F2}
\end{figure}

\subsection{Overall Results}

Results for the 12 stations over 731 verification days are shown in Table \ref{T1}. The stations in the table are ordered with respect to the average amount of precipitation totals during the considered time period, from 1.4577 at Artern to 5.5148 at Zugspitze. Originally, the 12 stations were selected from a larger data set containing 155 possible stations by their average precipitation totals. The first and the last three stations  admit lowest and largest average precipitation totals, respectively, while the remaining six stations scatter around a median of average precipitation totals.

As can be seen from Table \ref{T1}, Chi0  performs well with respect to the mean CRPS and does not produce unusual large CRPS values as the other two methods do (quite rarely though) for some of the stations, the reason for the latter not being quite clear. Of course, extreme outlying score values have a strong influence on the mean score. For example, if for station 'Berlin Tegel' the maximal CRPS is removed for method CSG0, then the maximum of the remaining values is 29.6 and the mean score is 0.7836, implying that Chi0 still admits smallest mean score value. We were not able to see any obvious reason nor pattern for the sporadic occurrence of unusual large score values for methods GEV0 and CSG0. Admittedly, due to optimization in connection with numerical integration, Chi0 is much slower than the other two. The assumed link (\ref{E3}) to the ensemble statistics $\overline{f}$ and $s_{f}$ obviously works well, and has been used with starting values $a=0.5$ and $b=c=d=1$.

The upper left diagram in Figure \ref{F2} shows a rank verification bar plot. It displays the frequency of the rank of the observation within the $m=50$ ensemble members over all 12 stations and $731$ verification days for each station.
From this, it may be concluded that the raw ensemble is badly calibrated revealing strong underdispersion. The other three diagrams show the PIT values of the three models for the same set of verification data. The PIT values are obtained by computing the predictive CDF at the verifying observation. However, since all three models have a point mass at 0, the usual PIT value is replaced by a random number from the interval between 0 and
the predictive CDF at the verifying observation whenever the verifying observation equals 0, see also \citet{baran2016censored}. Hence, when replicated the PIT histograms may slightly vary. From the diagrams, it is seen that all three models still admit some issues with underdispersion. As it seems, CGS0 and Chi0 perform better than GEV0, a fact which is also supported by the CRPS values from Table \ref{T1}. In general, CGS0 and Chi0 perform similar with a slight advantage on Chi0 as is also supported by Table \ref{T1}.

\begin{figure}[bt]
\centering
\includegraphics[width=38pc,angle=0]{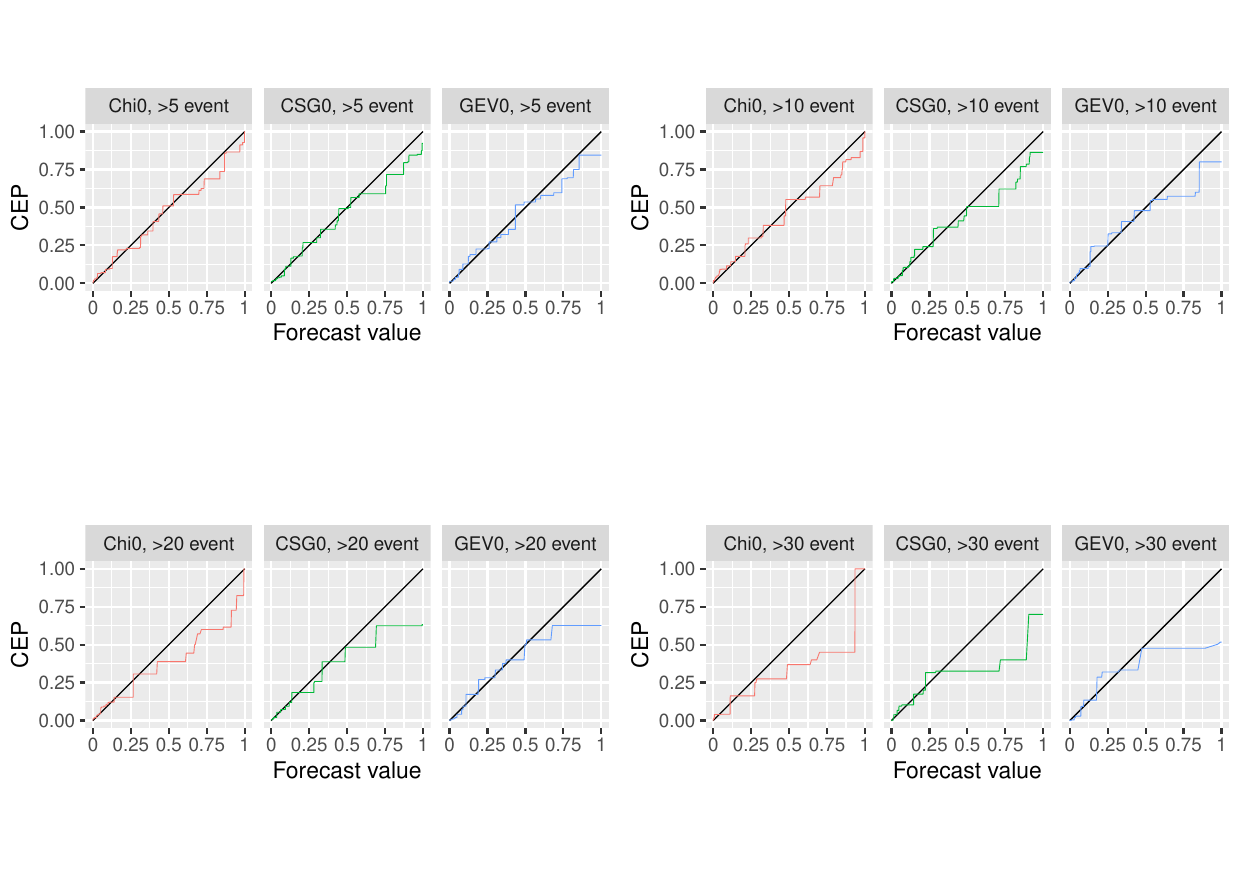}\\
  \caption{Reliability diagrams of the three considered methods for four precipitation events, 12 stations and $12 \times 713 = 8772$ verification days}\label{F4}
\end{figure}

\subsection{Precipitation Events}

In order to assess the forecast performance when predicting a binary event of the form that the observed precipitation amount $y$ exceeds a given threshold $\tau$, another proper scoring rule, the mean Brier Score \citep{brier1950verification, GneitingRaftery2007}, can be utilized:
\begin{equation}\label{E5}
\overline{\text{BS}} = \frac{1}{n}\sum_{i=1}^{n} (p_{i} -  y_{i})^2\; .
\end{equation}
Here, $n$ is the number of verification points, $y_{i}$ is the dichotomized observed event (either $0$ or $1$ with $1$ indicating the event), and $p_{i}$ is the forecast probability for the considered event obtained from the predictive probability distribution. The Brier Score for the raw ensemble may be obtained by the same formula (\ref{E5}) when $p_{i}$ is replaced by the relative frequency of the ensemble members predicting the event.

As the CRPS can be seen as the integral of the Brier Score over the continuum of all possible threshold values, it measures an overall forecast performance.
The Brier Score, on the contrary, can be used to assess the forecast accuracy for certain (e.g. higher) levels of precipitation (defined by the binary event). In principle, smaller values of $\overline{\text{BS}}$  indicate better forecast performance.
However, as the frequency of observed successes decreases with increasing thresholds, the corresponding $\overline{\text{BS}}$ values have entirely different magnitudes and cannot be directly compared.
In our analysis we consider the binary events '$>\tau$' for threshold values {$\tau= 5$}mm, {$10$}mm, {$20$}mm, and {$30$}mm.

From \citet{dimitriadis2021stable}, the mean Brier Score  can be decomposed as
\begin{equation}
\overline{\text{BS}} = \text{MCB}- \text{DSC} + \text{UNC}\; ,
\end{equation}
where $\text{MCB}$ is a measure of miscalibration (smaller values are better), $\text{DSC}$ is a measure of
discrimination (larger values are better), and $\text{UNC}$ is a measure of uncertainty. This may be seen as a generalization of the classical Brier Score decomposition by \citet{murphy1973new}, and may also be carried out for other proper scoring rules, see \citet{dimitriadis2021stable} for details and \citet{arnold2023decompositions} for a similar decomposition of the mean CRPS.

\begin{table}
\begin{tabular}{c|ccc|ccc}
 & MCB & DSC & UNC & MCB & DSC & UNC\\
 \hline
 & \multicolumn{3}{c|}{$ >5$mm} & \multicolumn{3}{c}{$ >10$mm}\\
 & \multicolumn{3}{c|}{($n_{e}=1380$)} & \multicolumn{3}{c}{($n_{e}=690$)}\\
\hline
Chi0 & 0.00157  & 0.0651 & 0.133 & 0.00114  & 0.0321 & 0.0725\\
CSG0 & 0.00206  & 0.0650 & 0.133 & 0.00170  & 0.0316 & 0.0725\\
GEV0 & 0.00275  & 0.0609 & 0.133 & 0.00235  & 0.0299 & 0.0725\\
\hline
 & \multicolumn{3}{c|}{$ >20$mm} & \multicolumn{3}{c}{$ >30$mm}\\
 & \multicolumn{3}{c|}{($n_{e}=201$)} & \multicolumn{3}{c}{($n_{e}=76$)}\\
\hline
Chi0 & 0.00100 & 0.00769 & 0.0224& 0.000778 & 0.00262 & 0.00859\\
CSG0 & 0.00172 & 0.00743 & 0.0224& 0.00104  & 0.00242 & 0.00859\\
GEV0 & 0.00187 & 0.00725 & 0.0224& 0.00116  & 0.00221 & 0.00859\\
\hline
\end{tabular}\caption{Decomposition of mean Brier Score for the three considered methods and four precipitation events, 12 stations and $12\times 713 = 8772$ verification days, where $n_{e}$ is the number of actually observed precipitation events}\label{T3}
\end{table}

Table \ref{T3} shows the corresponding values for the three methods Chi0, CSG0 and EVG0, computed with the {\sf R} package
{\tt reliabilitydiag}, see \citet{dimitriadis2021stable}. It is seen that for each of the four events, MCB of Chi0 is smallest, while DSC is largest. Furthermore, CSG0 performs better than GEV0 with respect to the two measures.

The forecast reliability may also be illustrated by so-called CORP reliability diagrams as displayed in Figure \ref{F4}. The diagrams
show estimated conditional event probabilities (CEP) plotted against forecast values. Horizontal segments can be interpreted as bins (intervals of predictive probabilities) with their positions on the $y$-axis indicating the bin-specific empirical event frequencies, see \citet{dimitriadis2021stable} for details. From this, together with the measures in Table \ref{T3}, one may say that the forecast performance specifically for moderate precipitation events does not strongly differ with respect to the three methods with a slight advantage laying on Chi0. Jointly with Figure \ref{F4} one may conclude that Chi0 provides the most noticeable improvement over CSG0 and GEV0 for larger forecast probabilities (roughly greater than 0.75), and for larger precipitation amounts (e.g. {$>20$}mm and {$>30$}mm).
This indicates that Chi0 is able to predict higher probabilities for an event more reliable than the distributions CSG0 and GEV0.

Since we are mainly interested in a comparison of the three EMOS methods, we did not provide a comparison with the raw ensemble performance here, but do so for a selected individual station in the  following subsection.

\begin{figure}[bt]
\centering
\includegraphics[width=34pc,angle=0]{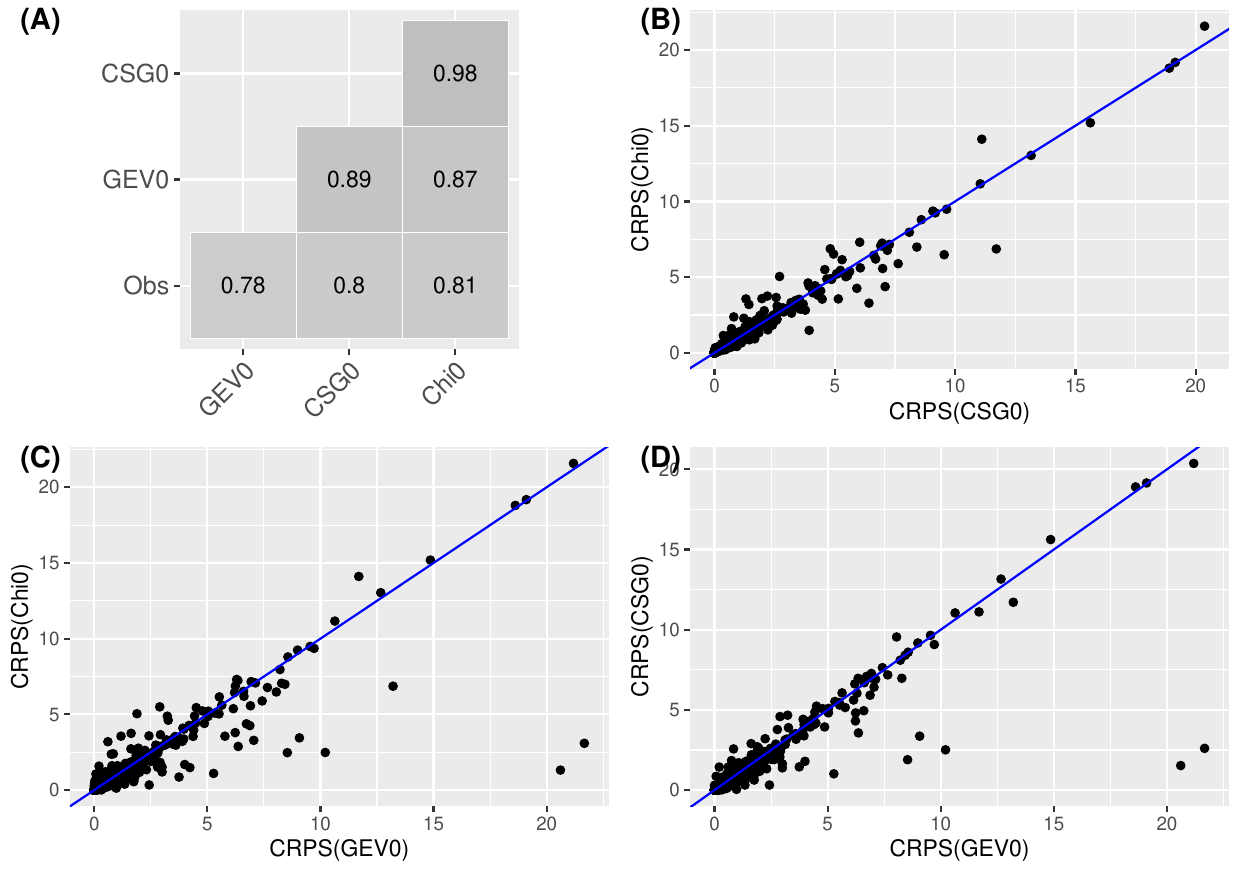}\\
  \caption{(A): Correlation  matrix of verifying precipitation observations (Obs) and CRPS values of the three models GEV0, CSG0,  and Chi0. (B), (C), (D): Scatter plots of CRPS pairs from the three models together with bisecting line for station Aachen  for 731 verification days}\label{F3}
\end{figure}

\subsection{Analysis of an Individual Station}\label{sec:onestation}

As noted above, CSG0 and Chi0 exhibit similar performance with a slight advantage on Chi0, while EVG0 is outperformed by the other two. This may also confirmed with an individual station,
but, of course, station specific characteristics can reveal a different behaviour in other cases. Here we consider the station 'Aachen' located in the western part of Germany.  For station 'Aachen', the accumulated 24h precipitation amounts are displayed in the first row of Figure \ref{F1}. In addition, the second row shows the corresponding forecast error $y- \overline{f}$ of the ensemble mean $\overline{f}$ as a point forecast for the observed precipitation $y$. The third row shows the corresponding CRPS values from the Chi0 model.

When considering a correlation matrix of the 731 CRPS values from the three models together with the corresponding precipitation observations one gets values displayed in Figure
\ref{F3} (A). It is seen that, on average, larger precipitation values come along with larger CRPS values. Moreover, a quite strong relationship between the performance of CSG0 and Chi0 may be concluded. Figure \ref{F3} (B), (C), and (D) also shows scatter plots with CRPS values of the three methods respectively paired. Points below the bisecting line indicate an advantage of the $y$-axis over the $x$-axis labeled method.

\begin{center}
\begin{table}[hbt]
\begin{tabular}{c|cccc}
& $ >2$mm  & $ >5$mm  & $ >10$mm  & $ >15$mm \\
& ($n_{e}=216$) & ($n_{e}=95$) & ($n_{e}=35$) & ($n_{e}=14$)\\
\hline
Ens  & 0.171 & 0.0857 & 0.0389 & 0.0127\\
Chi0 & 0.140 & 0.0893 & 0.0366 & 0.0129\\
CSG0 & 0.136 & 0.0890 & 0.0389 & 0.0137\\
GEV0 & 0.148 & 0.0969 & 0.0406 & 0.0155\\
\hline
\end{tabular}\caption{Mean Brier score  $\overline{\text{BS}}$ for station Aachen and 731 verification days for four precipitation events with $n_{e}$ respective event observations}\label{T4}
\end{table}
\end{center}

Table \ref{T4} shows a comparison of mean Brier Score values for four precipitation events with respect to the three considered postprocessing methods Chi0, CSG0, GEV0, as well as the raw ensemble (Ens). As already mentioned, the Brier score values of the considered events may not directly comparable as they are also based on different event sizes (indicated in brackets). However, our focus here is not the explicit comparison of the performance when predicting the different events, but the comparison of the different postprocessing models when predicting the same event.
For each of the four events, the order of magnitude of the score value for Ens appears to be in line with the corresponding score value for each of the three methods. For the two events '{$>5$ mm}' and '{$>15$ mm}' the raw ensemble even admits the smallest score. Moreover, Chi0 performs best in case '{$>10$ mm}' and second best in case '{$>15$ mm}'. In case of '{$>2$ mm}' method CSG0 performs best followed by Chi0.

\section*{Conclusion}

The above small study independently supports the conclusion from \citet{baran2016censored} that CSG0 has the potential to outperform EVG0. Moreover it has been shown that the zero degree of freedom non-central chi squared distribution (Chi0 EMOS model) also serves well as a predictive distribution in ensemble postprocessing of quantitative precipitation. The results can thus be seen as a motivation for further studying its properties and predictive performance, for postprocessing precipitation as well as weather quantities with similar properties.

\section*{Acknowledgements}

The authors gratefully acknowledge support by Deutsche Forschungsgemeinschaft (DFG) under grant number 395388010. Annette M\"{o}ller also gratefully acknowledges support by Deutsche Forschungsgemeinschaft (DFG) under grant number 520017589. Furthermore, we are grateful to the European Centre for Medium-Range Weather Forecasts (ECMWF) and the German Weather Service (DWD) for providing forecast and observation data, respectively.

\bibliography{ZeroDegree}

\begin{thebibliography}{50}
\providecommand{\natexlab}[1]{#1}
\providecommand{\url}[1]{\texttt{#1}}
\expandafter\ifx\csname urlstyle\endcsname\relax
  \providecommand{\doi}[1]{doi: #1}\else
  \providecommand{\doi}{doi: \begingroup \urlstyle{rm}\Url}\fi

\bibitem[Arnold et~al.(2023)Arnold, Walz, Ziegel, and
  Gneiting]{arnold2023decompositions}
S.~Arnold, E.-M. Walz, J.~Ziegel, and T.~Gneiting.
\newblock Decompositions of the mean continuous ranked probability score.
\newblock \emph{arXiv}, 2023.
\newblock URL \url{https://doi.org/10.48550/arXiv.2311.14122}.

\bibitem[Baran and M{\"o}ller(2015)]{baran2015joint}
S.~Baran and A.~M{\"o}ller.
\newblock Joint probabilistic forecasting of wind speed and temperature using
  {B}ayesian model averaging.
\newblock \emph{Environmetrics}, 26:\penalty0 120--132, 2015.

\bibitem[Baran and Nemoda(2016)]{baran2016censored}
S.~Baran and D.~Nemoda.
\newblock Censored and shifted gamma distribution based {EMOS} model for
  probabilistic quantitative precipitation forecasting.
\newblock \emph{Environmetrics}, 27\penalty0 (5):\penalty0 280--292, 2016.

\bibitem[Brier(1950)]{brier1950verification}
G.~W. Brier.
\newblock Verification of forecasts expressed in terms of probability.
\newblock \emph{Monthly Weather Review}, 78\penalty0 (1):\penalty0 1--3, 1950.

\bibitem[Buizza et~al.(2007)Buizza, Bidlot, Wedi, Fuentes, Hamrud, Holt, and
  Vitart]{buizza2007new}
R.~Buizza, J.-R. Bidlot, N.~Wedi, M.~Fuentes, M.~Hamrud, G.~Holt, and
  F.~Vitart.
\newblock The new {ECMWF} {VAREPS} (variable resolution ensemble prediction
  system).
\newblock \emph{Quarterly Journal of the Royal Meteorological Society},
  133\penalty0 (624):\penalty0 681--695, 2007.

\bibitem[Dawid(1984)]{Dawid1984}
A.~P. Dawid.
\newblock Present position and potential developments: Some personal views:
  Statistical theory: The prequential approach.
\newblock \emph{Journal of the Royal Statistical Society. Series A},
  147\penalty0 (2):\penalty0 278, 1984.

\bibitem[Dimitriadis et~al.(2021)Dimitriadis, Gneiting, and
  Jordan]{dimitriadis2021stable}
T.~Dimitriadis, T.~Gneiting, and A.~I. Jordan.
\newblock Stable reliability diagrams for probabilistic classifiers.
\newblock \emph{Proceedings of the National Academy of Sciences}, 118\penalty0
  (8):\penalty0 e2016191118, 2021.
\newblock URL \url{https://doi.org/10.1073/pnas.2016191118}.

\bibitem[Dunn(2004)]{dunn2004occurrence}
P.~K. Dunn.
\newblock Occurrence and quantity of precipitation can be modelled
  simultaneously.
\newblock \emph{International Journal of Climatology}, 24\penalty0
  (10):\penalty0 1231--1239, 2004.

\bibitem[Fraley et~al.(2022)Fraley, Raftery, Sloughter, Gneiting, and
  of~Washington.]{RPensembleBMA}
C.~Fraley, A.~E. Raftery, J.~M. Sloughter, T.~Gneiting, and U.~of~Washington.
\newblock \emph{ensembleBMA: Probabilistic Forecasting using Ensembles and
  Bayesian Model Averaging}, 2022.
\newblock URL \url{https://CRAN.R-project.org/package=ensembleBMA}.
\newblock R package version 5.1.8.

\bibitem[Gebetsberger et~al.(2016)Gebetsberger, Messner, Mayr, and
  Zeileis]{gebetsberger2016tricks}
M.~Gebetsberger, J.~W. Messner, G.~J. Mayr, and A.~Zeileis.
\newblock Tricks for improving non-homogeneous regression for probabilistic
  precipitation forecasts: Perfect predictions, heavy tails, and link
  functions.
\newblock Technical report, Working Papers in Economics and Statistics,
  University of Insbruck, 2016.

\bibitem[Gneiting and Katzfuss(2014)]{gneiting2014probabilistic}
T.~Gneiting and M.~Katzfuss.
\newblock Probabilistic forecasting.
\newblock \emph{Annual Review of Statistics and Its Application}, 1:\penalty0
  125--151, 2014.

\bibitem[Gneiting and Raftery(2005)]{GneitingRaftery2005}
T.~Gneiting and A.~Raftery.
\newblock Weather forecasting with ensemble methods.
\newblock \emph{Science}, 310:\penalty0 248--249, 2005.

\bibitem[Gneiting and Raftery(2007)]{GneitingRaftery2007}
T.~Gneiting and A.~E. Raftery.
\newblock Strictly proper scoring rules, prediction, and estimation.
\newblock \emph{Journal of the American Statistical Association}, 102:\penalty0
  359--378, 2007.

\bibitem[Gneiting et~al.(2005)Gneiting, Raftery, Westveld~III, and
  Goldman]{gneiting2005calibrated}
T.~Gneiting, A.~E. Raftery, A.~H. Westveld~III, and T.~Goldman.
\newblock Calibrated probabilistic forecasting using ensemble model output
  statistics and minimum {CRPS} estimation.
\newblock \emph{Monthly Weather Review}, 133:\penalty0 1098--1118, 2005.

\bibitem[Grimit et~al.(2006)Grimit, Gneiting, Berrocal, and
  Johnson]{grimit2006continuous}
E.~P. Grimit, T.~Gneiting, V.~J. Berrocal, and N.~A. Johnson.
\newblock The continuous ranked probability score for circular variables and
  its application to mesoscale forecast ensemble verification.
\newblock \emph{Quarterly Journal of the Royal Meteorological Society},
  132\penalty0 (621C):\penalty0 2925--2942, 2006.

\bibitem[Hasan and Dunn(2011)]{hasan2011two}
M.~M. Hasan and P.~K. Dunn.
\newblock Two {T}weedie distributions that are near-optimal for modelling
  monthly rainfall in {A}ustralia.
\newblock \emph{International Journal of Climatology}, 31\penalty0
  (9):\penalty0 1389--1397, 2011.

\bibitem[Hemri et~al.(2014)Hemri, Scheuerer, Pappenberger, Bogner, and
  Haiden]{hemri2014trends}
S.~Hemri, M.~Scheuerer, F.~Pappenberger, K.~Bogner, and T.~Haiden.
\newblock Trends in the predictive performance of raw ensemble weather
  forecasts.
\newblock \emph{Geophysical Research Letters}, 41\penalty0 (24):\penalty0
  9197--9205, 2014.

\bibitem[Hjort(1988)]{hjort1988eccentric}
N.~L. Hjort.
\newblock The eccentric part of the noncentral chi square.
\newblock \emph{The American Statistician}, 42\penalty0 (2):\penalty0 130--132,
  1988.

\bibitem[Javanshiri et~al.(2021)Javanshiri, Fathi, and
  Mohammadi]{javanshiri2021comparison}
Z.~Javanshiri, M.~Fathi, and S.~A. Mohammadi.
\newblock Comparison of the {BMA} and {EMOS} statistical methods for
  probabilistic quantitative precipitation forecasting.
\newblock \emph{Meteorological Applications}, 28\penalty0 (1):\penalty0 e1974,
  2021.
\newblock URL \url{https://doi.org/10.1002/met.1974}.

\bibitem[Jobst et~al.(2023{\natexlab{a}})Jobst, M{\"o}ller, and
  Gro{\ss}]{jobst2023d}
D.~Jobst, A.~M{\"o}ller, and J.~Gro{\ss}.
\newblock D-vine-copula-based postprocessing of wind speed ensemble forecasts.
\newblock \emph{Quarterly Journal of the Royal Meteorological Society},
  149\penalty0 (755):\penalty0 2575--2597, 2023{\natexlab{a}}.

\bibitem[Jobst et~al.(2023{\natexlab{b}})Jobst, M{\"o}ller, and
  Gro{\ss}]{jobst2023d2}
D.~Jobst, A.~M{\"o}ller, and J.~Gro{\ss}.
\newblock D-vine gam copula based quantile regression with application to
  ensemble postprocessing.
\newblock \emph{arXiv}, 2023{\natexlab{b}}.
\newblock URL \url{https://doi.org/10.48550/arXiv.2309.05603}.

\bibitem[Jobst et~al.(2024)Jobst, Möller, and Groß]{jobstetal2024}
D.~Jobst, A.~Möller, and J.~Groß.
\newblock Time series based ensemble model output statistics for temperature
  forecasts postprocessing.
\newblock \emph{arXiv}, 2024.
\newblock URL \url{https://doi.org/10.48550/arXiv.2402.00555}.

\bibitem[Johnson et~al.(1995)Johnson, Kotz, and
  Balakrishnan]{johnson1995continuous}
N.~L. Johnson, S.~Kotz, and N.~Balakrishnan.
\newblock \emph{Continuous {U}nivariate {D}istributions, {V}olume 2}.
\newblock John Wiley \& Sons, New York, 1995.

\bibitem[Jones(1987)]{jones1987relationship}
M.~Jones.
\newblock On the relationship between the {P}oisson-exponential model and the
  non-central chi-squared distribution.
\newblock \emph{Scandinavian Actuarial Journal}, 1987\penalty0 (1-2):\penalty0
  104--109, 1987.

\bibitem[Jordan et~al.(2019)Jordan, Kr\"uger, and Lerch]{jordan2017evaluating}
A.~Jordan, F.~Kr\"uger, and S.~Lerch.
\newblock Evaluating probabilistic forecasts with {scoringRules}.
\newblock \emph{Journal of Statistical Software}, 90\penalty0 (12):\penalty0
  1--37, 2019.
\newblock URL \url{https://doi.org/10.18637/jss.v090.i12}.

\bibitem[Khajehei and Moradkhani(2017)]{khajehei2017towards}
S.~Khajehei and H.~Moradkhani.
\newblock Towards an improved ensemble precipitation forecast: a probabilistic
  post-processing approach.
\newblock \emph{Journal of Hydrology}, 546:\penalty0 476--489, 2017.

\bibitem[Leutbecher and Palmer(2008)]{LeutbecherPalmer2008}
M.~Leutbecher and T.~N. Palmer.
\newblock Ensemble forecasting.
\newblock \emph{Journal of Computational Physics}, 227:\penalty0 3515--3539,
  2008.

\bibitem[M{\"o}ller and Gro{\ss}(2016)]{moller2016probabilistic}
A.~M{\"o}ller and J.~Gro{\ss}.
\newblock Probabilistic temperature forecasting based on an ensemble
  autoregressive modification.
\newblock \emph{Quarterly Journal of the Royal Meteorological Society},
  142:\penalty0 1385--1394, 2016.

\bibitem[M{\"o}ller and Gro{\ss}(2020)]{moller2020probabilistic}
A.~M{\"o}ller and J.~Gro{\ss}.
\newblock Probabilistic temperature forecasting with a heteroscedastic
  autoregressive ensemble postprocessing model.
\newblock \emph{Quarterly Journal of the Royal Meteorological Society},
  146\penalty0 (726):\penalty0 211--224, 2020.

\bibitem[M{\"o}ller et~al.(2013)M{\"o}ller, Lenkoski, and
  Thorarinsdottir]{moller2013multivariate}
A.~M{\"o}ller, A.~Lenkoski, and T.~L. Thorarinsdottir.
\newblock Multivariate probabilistic forecasting using ensemble {B}ayesian
  model averaging and copulas.
\newblock \emph{Quarterly Journal of the Royal Meteorological Society},
  139:\penalty0 982--991, 2013.

\bibitem[Molteni et~al.(1996)Molteni, Buizza, Palmer, and
  Petroliagis]{molteni1996ecmwf}
F.~Molteni, R.~Buizza, T.~N. Palmer, and T.~Petroliagis.
\newblock The {ECMWF} ensemble prediction system: Methodology and validation.
\newblock \emph{Quarterly Journal of the Royal Meteorological Society},
  122\penalty0 (529):\penalty0 73--119, 1996.

\bibitem[Murphy(1973)]{murphy1973new}
A.~H. Murphy.
\newblock A new vector partition of the probability score.
\newblock \emph{Journal of Applied Meteorology and Climatology}, 12\penalty0
  (4):\penalty0 595--600, 1973.

\bibitem[Palmer(2002)]{Palmer2002}
T.~Palmer.
\newblock The economic value of ensemble forecasts as a tool for risk
  assessment: From days to decades.
\newblock \emph{Quarterly Journal of the Royal Meteorological Society},
  128:\penalty0 747--774, 2002.

\bibitem[{R Core Team}(2024)]{Rsoftware}
{R Core Team}.
\newblock \emph{R: A Language and Environment for Statistical Computing}.
\newblock R Foundation for Statistical Computing, Vienna, Austria, 2024.
\newblock URL \url{https://www.R-project.org/}.

\bibitem[Raftery et~al.(2005)Raftery, Gneiting, Balabdaoui, and
  Polakowski]{Raftery&2005}
A.~Raftery, T.~Gneiting, F.~Balabdaoui, and M.~Polakowski.
\newblock Using {B}ayesian model averaging to calibrate forecast ensembles.
\newblock \emph{Monthly Weather Review}, 133:\penalty0 1155--1174, 2005.

\bibitem[Saedi et~al.(2020)Saedi, Saghafian, Moazami, and
  Aminyavari]{saedi2020performance}
A.~Saedi, B.~Saghafian, S.~Moazami, and S.~Aminyavari.
\newblock Performance evaluation of sub-daily ensemble precipitation forecasts.
\newblock \emph{Meteorological Applications}, 27\penalty0 (1):\penalty0 e1872,
  2020.
\newblock URL \url{https://doi.org/10.1002/met.1872}.

\bibitem[Scheuerer(2014)]{scheuerer2014probabilistic}
M.~Scheuerer.
\newblock Probabilistic quantitative precipitation forecasting using ensemble
  model output statistics.
\newblock \emph{Quarterly Journal of the Royal Meteorological Society},
  140\penalty0 (680):\penalty0 1086--1096, 2014.

\bibitem[Scheuerer and Hamill(2015)]{scheuerer2015statistical}
M.~Scheuerer and T.~M. Hamill.
\newblock Statistical postprocessing of ensemble precipitation forecasts by
  fitting censored, shifted gamma distributions.
\newblock \emph{Monthly Weather Review}, 143\penalty0 (11):\penalty0
  4578--4596, 2015.

\bibitem[Schlosser et~al.(2019)Schlosser, Hothorn, Stauffer, and
  Zeileis]{schlosser2019distributional}
L.~Schlosser, T.~Hothorn, R.~Stauffer, and A.~Zeileis.
\newblock Distributional regression forests for probabilistic precipitation
  forecasting in complex terrain.
\newblock \emph{The Annals of Applied Statistics}, 13\penalty0 (3), 2019.

\bibitem[Siegel(1979)]{siegel1979noncentral}
A.~F. Siegel.
\newblock The noncentral chi-squared distribution with zero degrees of freedom
  and testing for uniformity.
\newblock \emph{Biometrika}, 66\penalty0 (2):\penalty0 381--386, 1979.

\bibitem[Siegel(1985)]{siegel1985modelling}
A.~F. Siegel.
\newblock Modelling data containing exact zeroes using zero degrees of freedom.
\newblock \emph{Journal of the Royal Statistical Society: Series B
  (Methodological)}, 47\penalty0 (2):\penalty0 267--271, 1985.

\bibitem[Sloughter et~al.(2007)Sloughter, Raftery, Gneiting, and
  Fraley]{sloughter2007probabilistic}
J.~M.~L. Sloughter, A.~E. Raftery, T.~Gneiting, and C.~Fraley.
\newblock Probabilistic quantitative precipitation forecasting using {B}ayesian
  model averaging.
\newblock \emph{Monthly Weather Review}, 135\penalty0 (9):\penalty0 3209--3220,
  2007.

\bibitem[Stauffer et~al.(2017{\natexlab{a}})Stauffer, Mayr, Messner, Umlauf,
  and Zeileis]{stauffer2017spatio}
R.~Stauffer, G.~J. Mayr, J.~W. Messner, N.~Umlauf, and A.~Zeileis.
\newblock Spatio-temporal precipitation climatology over complex terrain using
  a censored additive regression model.
\newblock \emph{International Journal of Climatology}, 37\penalty0
  (7):\penalty0 3264--3275, 2017{\natexlab{a}}.
\newblock URL \url{https://doi.org/10.1002/joc.4913}.

\bibitem[Stauffer et~al.(2017{\natexlab{b}})Stauffer, Umlauf, Messner, Mayr,
  and Zeileis]{stauffer2017ensemble}
R.~Stauffer, N.~Umlauf, J.~W. Messner, G.~J. Mayr, and A.~Zeileis.
\newblock Ensemble postprocessing of daily precipitation sums over complex
  terrain using censored high-resolution standardized anomalies.
\newblock \emph{Monthly Weather Review}, 145\penalty0 (3):\penalty0 955--969,
  2017{\natexlab{b}}.

\bibitem[Talagrand et~al.(1997)Talagrand, Vautard, and Strauss]{Talagrand1997}
O.~Talagrand, R.~Vautard, and B.~Strauss.
\newblock Evaluation of probabilistic prediction systems.
\newblock \emph{Proc. Workshop on Predictability}, pages 1--25, 1997.
\newblock URL
  \url{https://www.ecmwf.int/en/elibrary/12555-evaluation-probabilistic-prediction-systems}.

\bibitem[Torgersen(1972)]{torgersen1972supplementary}
E.~N. Torgersen.
\newblock Supplementary notes on linear models.
\newblock \emph{Preprint series. Statistical Memoirs}, 1972.
\newblock URL \url{http://urn.nb.no/URN:NBN:no-58681}.

\bibitem[{van Buuren} and Groothuis-Oudshoorn(2011)]{miceRpackage}
S.~{van Buuren} and K.~Groothuis-Oudshoorn.
\newblock {mice}: Multivariate imputation by chained equations in {R}.
\newblock \emph{Journal of Statistical Software}, 45\penalty0 (3):\penalty0
  1--67, 2011.
\newblock URL \url{https://doi.org/10.18637/jss.v045.i03}.

\bibitem[Wilks and Hamill(2007)]{WilksHamill2007}
D.~Wilks and T.~Hamill.
\newblock Comparison of ensemble-{MOS} methods using {GFS} reforecasts.
\newblock \emph{Monthly Weather Review}, 135:\penalty0 2379--2390, 2007.

\bibitem[Wilks(2011)]{wilks2011statistical}
D.~S. Wilks.
\newblock \emph{Statistical {M}ethods in the {A}tmospheric {S}ciences}.
\newblock Academic Press, 2011.

\bibitem[Yuen et~al.(2018)Yuen, Baran, Fraley, Gneiting, Lerch, Scheuerer, and
  Thorarinsdottir]{RPensembleMOS}
R.~Yuen, S.~Baran, C.~Fraley, T.~Gneiting, S.~Lerch, M.~Scheuerer, and
  T.~Thorarinsdottir.
\newblock \emph{ensembleMOS: Ensemble Model Output Statistics}, 2018.
\newblock URL \url{https://CRAN.R-project.org/package=ensembleMOS}.
\newblock R package version 0.8.2.

\end{thebibliography}
\bibliographystyle{abbrvnat.bst}

\end{document}